\documentclass[prl,preprint,twocolumn,superscriptaddress,floatfix,showpacs,10pt]{revtex4-1}%
\usepackage{amsfonts,amsmath,amssymb,bm,bbm}
\usepackage{graphicx}
\usepackage{xcolor}
\allowdisplaybreaks
\usepackage[colorlinks=true,urlcolor=blue,anchorcolor=blue,citecolor=blue,filecolor=blue,linkcolor=blue,menucolor=blue,pagecolor=blue,linktocpage=true,pdfproducer=medialab,pdfa=true]{hyperref}
\usepackage{dsfont}
\usepackage{calrsfs}
\usepackage{footmisc}
\DeclareMathAlphabet{\pazocal}{OMS}{zplm}{m}{n}
\DeclareMathAlphabet{\mathdsl}{U}{bbm}{m}{sl}

\usepackage{braket}

\newcommand{\dd}{\mathrm{d}}
\newcommand{\F}{{\cal F}}

\newcommand{\h}{\lrcorner}
\newcommand{\T}{\mathbb{T}}

\makeatletter
\DeclareFontFamily{OMX}{MnSymbolE}{}
\DeclareSymbolFont{MnLargeSymbols}{OMX}{MnSymbolE}{m}{n}
\SetSymbolFont{MnLargeSymbols}{bold}{OMX}{MnSymbolE}{b}{n}
\DeclareFontShape{OMX}{MnSymbolE}{m}{n}{
    <-6>  MnSymbolE5
   <6-7>  MnSymbolE6
   <7-8>  MnSymbolE7
   <8-9>  MnSymbolE8
   <9-10> MnSymbolE9
  <10-12> MnSymbolE10
  <12->   MnSymbolE12
}{}
\DeclareFontShape{OMX}{MnSymbolE}{b}{n}{
    <-6>  MnSymbolE-Bold5
   <6-7>  MnSymbolE-Bold6
   <7-8>  MnSymbolE-Bold7
   <8-9>  MnSymbolE-Bold8
   <9-10> MnSymbolE-Bold9
  <10-12> MnSymbolE-Bold10
  <12->   MnSymbolE-Bold12
}{}

\let\llangle\@undefined
\let\rrangle\@undefined
\DeclareMathDelimiter{\llangle}{\mathopen}%
                     {MnLargeSymbols}{'164}{MnLargeSymbols}{'164}
\DeclareMathDelimiter{\rrangle}{\mathclose}%
                     {MnLargeSymbols}{'171}{MnLargeSymbols}{'171}               
\makeatother
\usepackage{stmaryrd}
\DeclareFontFamily{U}{mathx}{\hyphenchar\font45}
\DeclareFontShape{U}{mathx}{m}{n}{<-> mathx10}{}
\DeclareSymbolFont{mathx}{U}{mathx}{m}{n}
\DeclareMathAccent{\widebar}{0}{mathx}{"73}
 \newcommand*{\commawedge}{%
  \mathpunct{%
    \mathop{,}\limits^{\wedge}%
  }%
}
\DeclareMathAlphabet\mathbfcal{OMS}{cmsy}{b}{n}

\begin{document}

\title{An Algebraic Classification of Solution Generating Techniques}
\preprint{MI-HET-762}

\author{Riccardo Borsato}
\email{riccardo.borsato@usc.es}
\affiliation{Instituto Galego de F\'isica de Altas Enerx\'ias (IGFAE), Universidade de Santiago de Compostela, Spain}

\author{Sibylle Driezen}
\email{sib.driezen@gmail.com}
\affiliation{Instituto Galego de F\'isica de Altas Enerx\'ias (IGFAE), Universidade de Santiago de Compostela, Spain}

\author{Falk Hassler}
\email{falk@fhassler.de}
\affiliation{George P. \& Cynthia Woods Mitchell Institute for Fundamental Physics and Astronomy,\\
Texas A\&M University, College Station, TX 77843, USA}

\begin{abstract}
\vspace{7pt}
We consider a two-fold problem:  on the one hand, the classification of a family of solution-generating techniques in (modified) supergravity and, on the other hand, the classification of a family of canonical transformations of 2-dimensional $\sigma$-models  giving rise to integrable-preserving transformations. Assuming a generalised Scherk-Schwarz ansatz, in fact, the two problems admit essentially the same algebraic formulation, emerging from an underlying double Lie algebra $\mathfrak d$.
After presenting our derivation of the classification, we discuss in detail  the relation to modified supergravity and the additional conditions to recover  the standard (unmodified) supergravity. Starting from our master equation --- that encodes all the possible continuous deformations allowed in the family of solution-generating techniques --- we show that these are classified by the Lie algebra cohomologies $H^2(\mathfrak h,\mathbb R)$ and $H^3(\mathfrak h,\mathbb R)$ of the maximally isotropic subalgebra $\mathfrak h$ of the double Lie algebra $\mathfrak d$. {We illustrate our results with a non-trivial example, the bi-Yang-Baxter-Wess-Zumino model.}
\end{abstract}

\maketitle

In this Letter we address two questions that, under a certain assumption to be explained below, happen to share a common answer. On the one hand, we want to explore and classify solution-generating techniques (SGTs) in supergravity. In fact, given the huge number of  string vacua it is preferable, when possible, to have an organising principle to better understand the space of solutions. Moreover, knowing how to generate  new solutions when starting from a known one turns out to be very useful for concrete applications. For simplicity, in most of the paper we will focus on the common Neveu-Schwarz-Neveu-Schwarz (NSNS) sector of the closed string (super)gravity equations, and at the end we will say how the corresponding transformation rules of the Ramond-Ramond (RR) fields can be found. Let us remark that our SGTs remain such even when including the first $\alpha'$-corrections in the cases of the bosonic  and the heterotic string, and (trivially) the type II superstring.

On the other hand, we are also interested in exploring and classifying canonical transformations of 2-dimensional $\sigma$-models. Our motivation comes from the study of integrable models, given that canonical transformations preserve the classical worldsheet integrability.
Let us remark that models related by canonical transformations are not necessarily  the same: the relation may be a hidden duality (such as T-duality), or a more complicated one  leading to a different spectrum because of the incompatibility with the worldsheet boundary conditions (as in the case of non-abelian T-duality~\cite{delaOssa:1992vci,Gasperini:1993nz,Giveon:1993ai} and homogeneous Yang-Baxter deformations~\cite{Klimcik:2008eq,Delduc:2013qra,Kawaguchi:2014qwa}). Therefore, the canonical transformations that we  consider --- and that for simplicity we will still call SGTs --- are a useful tool to generate new integrable models when starting from known ones. 

In fact, one may even combine the two motivations and consider SGTs on \emph{integrable string} $\sigma$-models. This direction of research is particularly interesting also because of the underlying integrability of various instances of the AdS/CFT correspondence (see~\cite{Beisert:2010jr} for a standard review), and the consequent hope of constructing generalisations of AdS/CFT that are still tractable by exact methods, at least in principle. See~\cite{Matsumoto:2014gwa,vanTongeren:2015uha,Guica:2017mtd,deLeeuw:2021ufg} for some examples of different methods in this direction of research.

The families of SGTs that we  classify are obtained by demanding the invariance of two objects --- called ``generalised fluxes'' $\mathbfcal{F}_{\bm{ABC}}$ and  $\mathbfcal{F}_{\bm{A}}$.
In the flux formulation of Double Field Theory (DFT)~\footnote{DFT~\cite{Siegel:1993th,Hull:2009mi,Hohm:2010jy,Hohm:2010pp} was reformulated in the flux formulation in~\cite{Geissbuhler:2013uka}. We recommend the review~\cite{Aldazabal:2013sca}. Ramond-Ramond fields may also be included, see~\cite{Hohm:2011dv,Coimbra:2011nw,Jeon:2012kd}.} --- which for us will be just a convenient rewriting of (modified) supergravity --- the dynamical fields are the generalised vielbein $\bm{E_A{}^M}$ and the generalised dilaton $\bm{d}$.  The NSNS sector of the DFT equations of motion, as well as their first $\alpha'$-correction, are  written exclusively in terms of the above generalised fluxes  (constructed from the fundamental fields $\bm{E_A{}^M}$ and $\bm{d}$) and their flat derivatives \cite{Geissbuhler:2013uka,Baron:2017dvb}. While demanding the invariance of the generalised fluxes (and their flat derivatives) to ensure that we map solutions to solutions~\footnote{The invariance of the generalised fluxes is only a sufficient condition for a SGT, and there may be more general symmetries of the DFT equations of motion (see e.g.~\cite{Demulder:2019vvh,Hronek:2020skb}).}, at the same time we will  make sure that $\bm{E_A{}^M}$ and $\bm{d}$ change non-trivially.

Going back to the discussion on generic $\sigma$-models, when considering the Hamiltonian formulation, it is possible to choose the phase-space variables in such a way that the Poisson bracket is controlled by an object that is again parameterised as the generalised flux $\mathbfcal{F}_{\bm{ABC}}$ of DFT. We refer to~\cite{Osten:2019ayq,Borsato:2021gma} for recent papers that include a review of this point. In this context, in order to capture the family of SGTs corresponding to canonical transformations, it is therefore enough to impose the invariance of  $\mathbfcal{F}_{\bm{ABC}}$ only.

Let us remark that this mechanism of invariance of the generalised fluxes is precisely what is behind (non-abelian) T-duality~\cite{delaOssa:1992vci,Gasperini:1993nz,Giveon:1993ai}, Poisson-Lie T-duality~\cite{Klimcik:1995ux,Klimcik:1995dy} and homogeneous Yang-Baxter deformations~\cite{Kawaguchi:2014qwa}, {as was shown in a series of papers, see e.g.~\cite{Hassler:2017yza,Demulder:2018lmj,Sakatani:2019jgu,Catal-Ozer:2019hxw,Catal-Ozer:2019tmm,Borsato:2020bqo,Hassler:2020tvz,Borsato:2020wwk,Codina:2020yma}.}
Importantly, it is  convenient to formulate them in the language of  generalised Scherk-Schwarz (gSS) compactifications \cite{Aldazabal:2011nj,Geissbuhler:2011mx,Grana:2012rr}. With the aim of clarifying the set-up and introduce the notation, we will now give a brief review of what is needed from gSS. For more details on the notational conventions as well as on the construction and motivations appearing in this Letter, we refer to~\cite{Borsato:2021gma} where the same objectives were addressed by different methods. The merit of the present Letter is  to provide a \emph{complete} and \emph{algebraic} classification of the family of SGTs under study.

\vspace{12pt}

\paragraph*{Generalised Scherk-Schwarz set-up.}
Consider splitting the physical coordinates as $\bm{x^m}=\{\dot x^{\dot \mu},x^\mu\}$ with $\bm{m} = 0, \ldots, D-1$ and $\mu =1 , \ldots d \leq D$. The gSS ansatz involves factorising the background as
\begin{equation}\label{eq:gSS}
\bm{E_A{}^M} ( \bm{x}) = \bm{ \dot{E}_A{}^I} (\dot{x}) \bm{ U_I{}^M} (x) , \quad \bm{d} (\bm{x}) = \dot{d} (\dot{x}) + \lambda (x) \ .
\end{equation}
The dotted fields depend only on the ``\textit{spectator}'' coordinates $\dot{x}^{\dot{\mu}}$ which will not participate in the SGT. The fields ${\bm U}$ and $\lambda$, on the other hand, will non-trivially transform, and we demand that they depend only on the ``\textit{internal}'' coordinates $x^\mu$ spanning a physical space denoted as ${\cal M}$. For our purposes, we do not need to demand $\cal M$ to be compact. Using the gSS jargon, we will refer to ${\bm U}$ and $\lambda$ as ``\textit{twists}''.  Additionally, in the gSS ansatz $\bm{ U_I{}^M}$ is of block form, being just the identity in the spectator block and acting  non-trivially only in the internal directions where it coincides with the matrix $U_I{}^M$. Here $I = 1, \ldots , 2d$ are algebra indices (as we will argue shortly) and $M = 1, \ldots , 2d$ are curved indices that are the doubled version of $\mu$~\footnote{See appendix A of~\cite{Borsato:2021gma} for a recap on the notational conventions used in this Letter.}. Under this ansatz, the generalised fluxes take the form $\mathbfcal{F}_{\bm{ABC}} = \mathbfcal{\dot F}_{\bm{ABC}} + \bm{ \dot{E}_A}{}^I  \bm{ \dot{E}_B}{}^J  \bm{ \dot{E}_C}{}^K  \F_{IJK} $ and $\mathbfcal{F}_{\bm{A}} = \mathbfcal{\dot F}_{\bm{A}} + \bm{ \dot{E}_A}{}^I   \F_{I}$, where dotted fields only depend on spectators, and
\begin{equation} \label{eq:fluxdef}
  \begin{aligned}
    \F_{IJK} &= 3 U_{[I}{}^M \partial_M U_J{}^N U_{K]N} , \\
   \F_I &= - \partial_M U_I{}^M + 2 U_I{}^M \partial_M \lambda \ .
 \end{aligned}
\end{equation}
As in gSS compactifications, we further assume that the fluxes $\F_{IJK}$ and $\F_I$ are $x$-independent (constant). This is  a natural assumption for gauged DFT and consistent truncations on generalised parallelisable spaces. The Bianchi identities and the section condition of DFT then imply that  $\F_{IJ}{}^K$ satisfy the Jacobi identity. Here, the algebra indices $I,J$ are raised and lowered by the constant metric $\eta_{IJ} = U_I{}^M \eta_{MN} U_J{}^N$,  with $\eta_{MN}$ the standard O$(d,d)$ metric. Therefore, $\F_{IJ}{}^K$ may be interpreted as the structure constants of a (doubled) Lie algebra ${\mathfrak d}$ with generators $T_I$, so that $[T_I, T_J] = \F_{IJ}{}^K T_K$. The metric $\eta_{IJ}$ is ad-invariant since $\F_{IJK}$ is completely antisymmetric. As we are about to explain, this algebraic interpretation allows one to construct the twists as well as the SGTs relating them in full generality. {Under the above assumptions}, the remaining Bianchi identities reduce to 
$\F_{IJ}{}^K \F_K = 0$ and $\F_{IJK} \F^{IJK} = 6 \F_I \F^I$.

\vspace{12pt}

\paragraph*{Parametrisation of the twist $U$.}  
As a consequence of the existence of a doubled Lie algebra $\mathfrak{d}$, the \textit{doubled} internal manifold (spanned by the coordinates $x^\mu$ together with their duals $\tilde{x}_{\tilde{\mu}}$) can be interpreted as a Lie group manifold that we call $D$. Solving the section condition on group manifolds \cite{Hassler:2016srl} requires the existence of a maximally isotropic subalgebra $\mathfrak{h} \subset \mathfrak{d}$. If we split the generators of $\mathfrak{d}$ as $T_I = \{T^i , T_i \}$, with $i = 1, \ldots , d$, and we denote the generators of $\mathfrak{h}$ by $T^i$, then  maximal isotropy of $\mathfrak h$ means that the metric $\eta_{IJ}$  takes the generic form
\begin{equation}\label{eq:noncanonicalmetric}
 \eta_{IJ}=\llangle T_I , T_J \rrangle =\left(
\begin{array}{cc}
0 & \eta^i{}_j\\
\eta_i{}^j& \eta_{ij}
\end{array}
\right),
\end{equation}
where $\eta^i{}_j =  \eta_j{}^i$. Moreover, the \emph{physical} internal space ${\cal M}$ is identified with the (right) coset ${\cal M} = H \backslash D$ \cite{Hassler:2016srl}. By using the group structure of $D$, this simple fact is enough to derive the twist $U_I{}^M$ satisfying \eqref{eq:fluxdef}, after a generic constant flux background $\F_{IJK}$ is given  \cite{Hassler:2019wvn} (for a comprehensive review see \cite{Demulder:2019vvh}). To be more precise, writing $U_I = k_I{}^\mu \partial_\mu + \phi_{I \mu} \dd x^\mu$ as a generalised vector and introducing a coset representative $m$ of ${\cal M} = H \backslash D$, the solution for the vector part is $k_I = M_I{}^j v_j{}^\mu \partial_\mu$, where $M_I{}^J T_J = m T_I m^{-1}$ and $v_i$ is such that $\iota_{v_i} dm m^{-1} = T_i$. The  solution for the one-form part is
$\phi_I =  \llangle M_I{}^J T_J - \frac{1}{2}\iota_{k_I} \dd m m^{-1}, \dd m m^{-1} \rrangle +\iota_{k_I} \Omega^{(2)} $,
where $\Omega^{(2)}$ is a two-form satisfying $\dd \Omega^{(2)} = -\frac{1}{6} \llangle \dd m m^{-1} \commawedge \dd m m^{-1} \wedge \dd m m^{-1} \rrangle$. Notice that it may not be possible to find a \emph{global} expression for $\Omega^{(2)}$, and that this object is defined only up to closed two-forms. When expanding the one-form $\dd m m^{-1}$ as
$
\dd m m^{-1} = ( v_\mu{}^i T_i + A_{\mu i} T^i ) \dd x^\mu 
$,
the resulting twist  in matrix form  reads as $U_I{}^M = M_I{}^J V_J{}^M$, with
\begin{equation}\label{eq:V-par}
V_I{}^M =   \begin{pmatrix}
\eta_j{}^i v_\mu{}^j & 0 \\ (A_{\nu k} \eta_{[i}{}^k v_{j]}{}^\nu  +\Omega^{(2)}_{ij} + \frac{1}{2} \eta_{ij} ) v_\mu{}^j & v_i{}^\mu
\end{pmatrix} ,
\end{equation}
where $\Omega^{(2)}_{ij} = v_i{}^\mu \Omega^{(2)}_{\mu\nu} v_j{}^\nu$ for $\Omega^{(2)} = \frac{1}{2} \Omega^{(2)}_{\mu\nu} \dd x^\mu \wedge \dd x^\nu$. Here we are denoting by $v_i{}^\mu$  the inverse of $v_\mu{}^i$.

The parameterisation of the twist $U_I{}^M$ given by equation~\eqref{eq:V-par} is the starting point of  our discussion. In fact, we first simplify the above solution by noticing that there always exists a $GL(2d)$ transformation $S$ that brings $\eta_{IJ}$ to the \emph{canonical} form of the $O(d,d)$ metric, while leaving the generalised frame $\bm{E_A{}^M}$ (and therefore the physical background) invariant. Firstly, notice that transforming the generators as $T'_{I'} = S_{I'}{}^I T_I$ with
\begin{equation} \label{eq:transformationcanonical}
  S_{I'}{}^I = \begin{pmatrix}
    s^{i'}{}_i & 0 \\ -\frac{1}{2} \eta_{i' j} s^j{}_i & \delta_{i'}{}^i
  \end{pmatrix} , ~~~ \text{and} ~~~ s^i{}_k \eta^k{}_j = \delta^i{}_j ,
\end{equation}
brings $\eta_{IJ}$, as anticipated, to the canonical form
\begin{equation} \label{eq:canonicalmetric}
\eta'_{I'J'} = S_{I'}{}^I \eta_{IJ} S_{J'}{}^J = \begin{pmatrix}
0 & \delta^{i'}{}_{j'} \\ \delta_{i'}{}^{j'} & 0
\end{pmatrix} .
\end{equation}
Secondly, since the transformation $S$ in~\eqref{eq:transformationcanonical} leaves the subalgebra $\mathfrak{h}$ invariant, we observe that we do not need to change the coset representative $m$ in the parameterisation. In components, we then have $M'_{I'}{}^{J'} = S_{I'}{}^I M_I{}^J (S^{-1})_J{}^{J'}$ and $v'_\mu{}^{i'} = v_\mu{}^{i} \delta_i{}^{i'}$, which straightforwardly lead to $k'_{I'} = M'_{I'}{}^{j'} v_{j'} = S_{I'}{}^I M_I{}^j v_j = S_{I'}{}^I  k_I $ and $\phi'_{I'} = S_{I'}{}^I  \phi_I $. Having shown that the twist transforms as $U'_{I'}{}^M = S_{I'}{}^I U_I{}^M$, if we also take into account that $S_{I'}{}^I$ is a constant matrix, we conclude that the effect of the latter is reabsorbed by the corresponding transformation of  the spectator field $\bm{\dot{E}_A{}^I}$, that is induced by the same change of basis. Therefore, the generalised frame $\bm{E_A{}^M}$ indeed remains invariant under this $S \in GL(2d)$. In other words, given a $\bm{E_A{}^M}$, we can always assume that the splitting in the gSS ansatz of~\eqref{eq:gSS} uses a basis for $\mathfrak d$ such that the metric $\eta_{IJ}$ is  canonical as in~\eqref{eq:canonicalmetric}. The expression for the twist then simplifies to
\begin{equation} \label{eq:canonicaltwist}
\boxed{U_{I}{}^M =M_{I}{}^{J} \begin{pmatrix}
 \delta^{j}{}_k & 0 \\ A_{\nu [j}  v_{k]}{}^\nu  +\Omega^{(2)}_{jk}   & \delta_{j}{}^k
\end{pmatrix} 
\begin{pmatrix}
 v_\mu{}^{k} & 0 \\ 0 & v_{k}{}^\mu
\end{pmatrix}}\,,
\end{equation}
where we omit the primes for brevity 
\footnote{{The expression \eqref{eq:canonicaltwist} for the twist $U$ already appeared in \cite{Demulder:2018lmj} but note that, here, we did not need to make their assumption (3.22). Hence, we find the same result in a more general setting.}}.
Note that an implicit freedom in this solution is to send $U_I{}^M\to W_I{}^JU_J{}^M$ with $W_I{}^J$  an automorphism of $\mathfrak d$. This fact will play a role later. 
Once we specify $\mathfrak{d}$ and thus  the  flux ${\cal F}_{IJK}$ --- whose components  are traditionally denoted as $\mathcal F_{ijk} = H_{ijk}$, $\mathcal F_{ij}{}^k = F_{ij}{}^k$, $\mathcal F_i{}^{jk} = Q_i{}^{jk}$ and $\mathcal F^{ijk} = R^{ijk}$ ---   we can construct the most generic twist by the formula \eqref{eq:canonicaltwist} above. Recall that in our choice of basis $T^i$ generates $\mathfrak{h}$, and therefore we always have $R^{ijk}=0$. Generic $O(d,d)$ transformations may spoil this choice.

\vspace{12pt}

\paragraph*{Parametrisation of the twist $\lambda$ and modified SUGRA.}  
 Using the expression \eqref{eq:canonicaltwist} for $U_I{}^M$, we can now integrate ${\cal F}_I$ to derive also a formula for the twist $\lambda$. While it is always possible to extract $Y_M =  \partial_M \lambda$ from \eqref{eq:fluxdef}, one has to additionally impose  the integrability condition $\partial_{[M} Y_{N]} = 0$ to permit a local solution for $\lambda$. 
When expressing the flux \eqref{eq:fluxdef} in terms of $Y_M$ only --- and thereby relaxing the identification $Y_M = \partial_M \lambda$ --- we find that 
 \footnote{This expression is derived from the Bianchi identity $D^K {\cal F}_{KIJ} + 2 D_{[I} {\cal F}_{J]} - {\cal F}^K {\cal F}_{KIJ} = {\cal Z}_{IJ}$ with $D_I \equiv U_I{}^M \partial_M$ and ${\cal Z}_{IJ} = (\partial^M \partial_M U_{[I}{}^N) U_{J]N} -  Y^M (\partial_M U_I{}^N) U_{JN} + 2 U_{[I}{}^M U_{J]}{}^N  \partial_M Y_N$ obtained by using the expression of ${\cal F}_I$ in terms of $Y_M$. Notice that the first term vanishes because of the section condition for $U_I{}^M$ and that the last term  would vanish if we identified $Y_M = \partial_M \lambda$. Using the expression \eqref{eq:canonicaltwist} and assuming constant fluxes and $\mathcal F_{IJ}{}^K \mathcal F_K = 0$, which in turn implies $M_I{}^J {\cal F}_J = {\cal F}_I$, we have ${\cal F}^i = Q_j{}^{ji} + v_
\mu{}^i Y^\mu$. 
}
\begin{equation}\label{eqn:intcondY}
 \partial_{[M} Y_{N]} =- \frac12 K^\mu \partial_\mu U^I{}_M U_{IN},
\end{equation}
where $K^\mu = Y^\mu= \frac12 (\mathcal{F}^i - Q_j{}^{ji}) v_i{}^\mu$. If $K^\mu$ vanishes, this integrability condition holds automatically. Otherwise,  without loss of generality, we may go to adapted coordinates $x^\mu = (y, \hat{x}^{\hat\mu})$  and take the  coset representative $m = \exp ( y K^i T_i ) n(\hat{x})$, with $K^i v_i{}^\mu = K^\mu$, to obtain $K^\mu \partial_\mu = \partial_y$. Then, imposing ${\cal F}_{[IJ}{}^L {\cal F}_{K]L}{}^M =0$ and ${\cal F}^K {\cal F}_{IJK} = 0$,   we observe that $U_I{}^M (y,\hat{x}) = U_I{}^M (y+a, \hat{x})$ for a constant shift $a$. Therefore,  $U_I{}^M$ does not depend on the coordinate $y$, and the right-hand-side of \eqref{eqn:intcondY} vanishes again. 
Lastly, 
 imposing also $\F_{IJK} \F^{IJK} = 6 \F_I \F^I$ we have the constraint $Y_M Y^M = \partial_M Y^M $. 
 
 Let us now parametrise $Y_M  =\partial_M \lambda= \begin{pmatrix}
K^\mu & \partial_\mu \lambda
\end{pmatrix}$ and point out that $Y_M$ must be independent of the dual coordinates, and that $\partial_M Y^M = 0$.  Notice that $Y_M Y^M = 0$ then implies that also $\lambda$ is $y$-independent, and consequently all the ordinary fields (metric, B-field and dilaton) will  be as well. Therefore, $K^\mu$ is a Killing vector of the background.
As a bonus, when combining all the above observations, we see that the twist $\lambda$ would still solve the section condition even when having a linear dependence on the dual $\tilde y$ coordinate. In particular, for $K^\mu \neq 0$, we have in adapted coordinates  $\lambda = \tilde{y} + \hat{\lambda}(\hat{x})$ for some function $\hat{\lambda}$ of the physical coordinates $\hat{x}$. 
When $K^\mu \neq 0$ we can view this as a $D$-dimensional background satisfying the modified SUGRA (mSUGRA)	 equations
 \cite{Arutyunov:2015mqj,Wulff:2016tju}, that depend explicitly on the vector $K^\mu$ and that reduce to the standard SUGRA equations when $K^\mu =0$.

Going back to the solution for the scalar twist, by computing ${\cal F}_i$
  we can now extract 
\begin{equation} \label{eq:lambdadiffeq}
  \boxed{\partial_\mu {\lambda}  = \frac12  ( v_\mu{}^i {\cal F}_i + A_{\mu i} Q_j{}^{ji} ) -  \frac{1}{2} \partial_\mu \log v +  K^\nu  \widetilde{B}_{\nu\mu} }\,,
\end{equation}
where $v = \det v_\mu{}^i$ 
 and
 $\widetilde{B}_{\mu\nu} =  v_\mu{}^i  (  A_{\rho [i}  v_{j]}{}^\rho  +\Omega^{(2)}_{ij}  ) v_\nu{}^j$. 
 This is a differential equation whose integration gives the solution for $\lambda$, which is therefore uniquely identified up to an inconsequential constant shift.
 
 The mSUGRA equations were already reinterpreted in the doubled formulation in \cite{Sakatani:2016fvh,Baguet:2016prz}.
To match with our discussion above, let us introduce the generalised vector $X_M \equiv  Y_M -  \partial_M ( \phi - 1/4 \log g)$, where $g$ is the determinant of the internal target-space metric and $\phi$ will be interpreted as the ordinary (internal) dilaton. Both  depend only on the physical coordinates $\hat{x}$. It is easy to see that $X_M$ satisfies two important properties: $\partial_{[M} X_{N]} = 0$ and $X_M X^M = 0$. 
The former implies
\begin{equation}\label{eqn:isomHMN}
  \mathcal L_X \mathcal H_{MN} = 0 ,
\end{equation}
because $X_M$ is a total derivative and thus mediates a trivial gauge transformation. In addition, in the adapted coordinates it is easy to show that
\begin{equation}
  \mathcal L_X \lambda = X^M \partial_M \lambda - \frac12 \partial_M X^M= 0 .
\end{equation}
 The above equations complete the ingredients required to relate to the modified DFT (mDFT) construction of  \cite{Sakatani:2016fvh}.
There $X_M$ was parametrised as
\begin{equation}
  X_M = \begin{pmatrix} K^\mu & Z_{\mu} - \partial_\mu \phi - K^\nu B_{\nu\mu} \end{pmatrix}\,,
\end{equation}
in terms of the vector field $K^\mu$, the (internal) target space $B$-field $B_{\mu\nu}$ and a one-form $Z_\mu$.
We remind that, when choosing  a gauge  such that $K^\mu$ is a Killing vector for the B-field \cite{Wulff:2016tju,Sakatani:2016fvh},  one has $X_M = \begin{pmatrix}
K^\mu & 0
\end{pmatrix}$. This agrees with our previous discussion by taking $\hat{\lambda} = \phi - 1/4 \log g$.

\vspace{12pt}

\paragraph*{Solution-generating techniques.} 
Let us now classify  SGTs that  leave the generalised fluxes $\mathbfcal{F}_{\bm{ABC}}$ and $\mathbfcal{F}_{\bm{A}}$ invariant, while changing non-trivially the twists $U$ and $\lambda$. We will assume that the generalised fluxes constructed out of the twists $U_I{}^M$ and $\lambda$  transform by a constant matrix that can be reabsorbed into the spectator field $\bm{\dot{E}_A{}^I}$.
More precisely, the condition to have a SGT will be 
\begin{equation} \label{eq:SGT}
\begin{aligned}
\mathcal F'_{I'J'K'} &=S_{I'}{}^IS_{J'}{}^J S_{K'}{}^K\mathcal F_{IJK},\quad \mathcal F'_{I'}=S_{I'}{}^I\mathcal F_{I}, \\
 \bm{\dot{E}'_A{}^{I'}} &=\bm{\dot{E}_A{}^J} \bm{(S^{-1})_J{}^{I'}},
\end{aligned}
\end{equation}
where $S_I{}^J$ is constant, and $\bm{(S^{-1})_J{}^I}$ is  non-trivial only in the internal directions where it acts as $(S^{-1})_J{}^I$.
 Importantly, we must mod out by those maps such that $U'_{I'}{}^M = S_{I'}{}^J U_J{}^M$, as these would leave $\bm{E_A{}^M}$ invariant and would therefore correspond to trivial SGTs.

Under the  transformation~\eqref{eq:SGT}, $\mathcal F'_{I'J'}{}^{K'}$ are still the structure constants of $\mathfrak d$, just expressed in a new basis $T'_{I'}=S_{I'}{}^JT_J$. As a consequence of the section condition, we still need to have a maximally isotropic subalgebra $\mathfrak h'\subset\mathfrak d$, possibly different from $\mathfrak h$, now spanned by the generators $T'^{i'}$. Following the previous discussion, we can safely assume that also in this new basis $\eta'_{I'J'}$ still takes the canonical form, and we conclude that $S_I{}^J\in O(d,d)\subset GL(2d)$. 
We parameterise it as~\footnote{This is the most generic $O(d,d)$ parameterisation up to subtleties discussed in appendix C of~\cite{Borsato:2021gma}.} $S=S^{(\rho)}S^{(b)}S^{(\beta)}T^n$ where the exponent $n$ can take the values $n=0,1$ and
\begin{equation}\label{eq:par-S}
\begin{aligned}
T_{I}{}^J&= 
\left(
\begin{array}{cc}
 0 & \delta^{ij}\\
\delta_{ij} & 0
\end{array}
\right),
\qquad
S^{(\rho)}_{I}{}^J=
\left(
\begin{array}{cc}
(\rho^t)^{i}{}_j & 0\\
0 & (\rho^{-1})_{i}{}^j
\end{array}
\right),\\
S^{(b)}_{I}{}^J&=
\left(
\begin{array}{cc}
\delta^i{}_j & 0\\
b_{ij} & \delta_i{}^j
\end{array}
\right),
\qquad
S^{(\beta)}_{I}{}^J=
\left(
\begin{array}{cc}
\delta^i{}_j & \beta^{ij}\\
0 & \delta_i{}^j
\end{array}
\right),
\end{aligned}
\end{equation}
with $\rho_i{}^j\in GL(d)$ and $b_{ij}=-b_{ji}, \beta^{ij}=-\beta^{ji}$.
Let us now analyse the consequences of each factor individually.

When the basis transformation is just $T'_{I'}=S^{(\rho)}_{I'}{}^IT_I$, the maximally isotropic subalgebra remains the same $(\mathfrak{h}'=\mathfrak h)$, and we are merely implementing a change of basis  that does not mix $T_i$ and $T^i$. Similarly to the discussion done to arrive at~\eqref{eq:canonicaltwist},  we keep the same $m\in \mathcal M$ as our coset representative and we easily  conclude that    the relation between the twists is just $U'_{I'}{}^M=S^{(\rho)}_{I'}{}^IU_I{}^M$. When the twists $U$ and $U'$ are related by the same transformation implementing the change of basis, we call this a \emph{rigid} $O(d,d)$ transformation. We  do not view it as a SGT because it would lead to $\bm{E'_A{}^M}=\bm{E_A{}^M}$. From now on we therefore set $\rho_i{}^j=\delta_i{}^j$ in~\eqref{eq:par-S}.
Similarly, also when $T'_{I'}=S^{(b)}_{I'}{}^IT_I$ the maximally isotropic subalgebra $\mathfrak h$ does not change, since $T'^{i'}=\delta^{i'}{}_i T^i$. We can again keep the same $m$, and check that also in this case we do not have a SGT because $U'_{I'}{}^M=S^{(b)}_{I'}{}^IU_I{}^M$, which prompts us to set also $b_{ij}=0$ in~\eqref{eq:par-S}.

Non-trivial SGTs are therefore  parameterised by $\beta^{ij}$, together with a possible additional action of the duality matrix $T$ when choosing the exponent $n=1$. These are in fact the transformations that change the choice of the maximally isotropic subalgebra within $\mathfrak{d}$. After the transformation, the original group element $m$ is not a  coset representative of $H'\backslash D$, and $U'_{I'}{}^M\neq S_{I'}{}^IU_I{}^M$.

Let us first set $n=0$. We must impose that the new generators $T'^{i}=\delta^{i}{}_jT^i+\beta^{ij}T_j$ still span a subalgebra $\mathfrak{h}'$ of $\mathfrak d$ (i.e.~$R'^{ijk} = 0$)~\footnote{If initially $H_{ijk}\neq0$ and $F_{ij}{}^k\neq 0$, the new generalised fluxes will be different from the original ones 
\begin{equation}\label{eqn:trHFQ}
\begin{aligned} 
H'_{ijk}={}&H_{ijk},
\quad F'_{ij}{}^k= F_{ij}{}^k- H_{ijl}\beta^{lk},\  \\
Q'_i{}^{jk} ={}&  Q_i{}^{jk} - 2  F_{il}{}^{[j} \beta^{k]l}  +   H_{imn} \beta^{jm}\beta^{kn}  .
\end{aligned}
\end{equation}
Nevertheless, we can always use a compensating  rigid $O(d,d)$ to redefine $\bm{\dot{E}''_A{}^I} \equiv \bm{\dot{E}'_A{}^J} \bm{S^{(\beta)}_J{}^I}= \bm{\dot{E}_A{}^I}$ and $U''_I{}^M\equiv (S^{(\beta)-1})_I{}^J U'_J{}^M$, so that $\mathcal F''_{IJK}= \mathcal F_{IJK}$.
}. This yields the condition
\begin{equation}\label{eqn:mastereq}
\boxed{
3 Q_l{}^{[ij}\beta^{k]l}
+3\beta^{l[i} F_{lm}{}^j\beta^{k]m}+\beta^{il}\beta^{jm}\beta^{kn}H_{lmn} = 0
}\,.
\end{equation}
In the above formula, the familiar reader may already recognise the 2-cocycle condition (when only $Q_i{}^{jk}\neq 0$) or the classical Yang-Baxter equation (when only $F_{ij}{}^k\neq 0$).
The above master equation will be further analysed later, and it will play an important role for a more explicit classification of the allowed deformations.
Having found $\mathcal F'_{I'J'K'} =S^{(\beta)}_{I'}{}^IS^{(\beta)}_{J'}{}^J S^{(\beta)}_{K'}{}^K\mathcal F_{IJK}$, we have a SGT in (m)SUGRA if also $\mathcal F'_{I'} =S^{(\beta)}_{I'}{}^I\mathcal F_I$ holds. First, notice that the combination of these two transformations leaves the conditions ${\cal F}_{[IJ}{}^L {\cal F}_{K]L}{}^M = 0$, ${\cal F}_{IJ}{}^{K} {\cal F}_K = 0 $ and ${\cal F}_{IJK}{\cal F}^{IJK} = 6 {\cal F}_I {\cal F}^I$ invariant. The SUGRA condition $K^\mu=0$, on the other hand, is preserved only if one imposes the following additional constraint 
\begin{equation}\label{eq:sugra-cond-beta}
\boxed{
  \beta^{ij} ( {\cal F}_j - F_{jk}{}^k ) - \beta^{jk} F_{jk}{}^i - H_{jmn} \beta^{jm} \beta^{in} = 0 }\,.
\end{equation}
The above equation can be interpreted as a supergravity condition on $\beta$, and it reduces to the ``unimodularity condition'' of~\cite{Borsato:2016ose} when specifying to the case of homogeneous Yang-Baxter deformations of isometric backgrounds. 

Let us now discuss the case $n=1$. The mutual presence of $S^{(\beta)}$ and $T$ can be understood as a composition of transformations. Therefore, once the classification of the $\beta$-deformations in the $n=0$ case is done, to have a full list of SGTs we only need to add those induced by the matrix $T$ alone. Under this transformation, the fluxes are related as $H_{ijk}\leftrightarrow R^{ijk}$ and $F_{ij}{}^k\leftrightarrow Q_i{}^{jk}$. Obviously, such duality maps exist only if  --- up to a rigid $O(d,d)$ transformation --- $R'^{ijk}=0$. 
Depending on which of the initial fluxes are non-vanishing, the  transformations induced by $T$ have been given different names in the literature: non-abelian T-duality (when initially only $F_{ij}{}^k\neq 0$) and Poisson-Lie T-duality (when both $F_{ij}{}^k\neq 0$ and $Q_i{}^{jk}\neq 0$). 
Also in this case of (generalised) dualities induced by $T$, the analysis of  $\mathcal F_I$ leads to a SUGRA condition. In particular, imposing $\mathcal F'_{I'}=T_{I'}{}^I\mathcal F_I$ implies firstly $\mathcal F'_i=\delta_{ij}\mathcal F^i$ (which may be interpreted as a differential equation for $\lambda'$) and secondly the SUGRA condition 
\begin{equation}\label{eq:sugra-cond-T}
\boxed{
F_{ji}{}^j=\mathcal F_i
}\,.\end{equation} 
Notice that in the case of isometric backgrounds one must have $\mathcal F_i=0$, and then the above condition reduces to the unimodularity condition of~\cite{Alvarez:1994np,Elitzur:1994ri}.

Let us remark that  when the SUGRA conditions~\eqref{eq:sugra-cond-beta} or~\eqref{eq:sugra-cond-T} are not satisfied, the SGT will relate a SUGRA  to a mSUGRA background, for which one can determine $K'^\mu$ accordingly.  In other words, the $\beta$-shifts and $T$-transformations that we consider are automatically  SGTs within mSUGRA/mDFT. 

To conclude the discussion on SGTs, we remind that in general we also have the freedom to change $\Omega^{(2)}$ in~\eqref{eq:canonicaltwist} by a closed two-form~\footnote{This closed two-form must not be exact, to avoid generating simple gauge transformations \cite{Hohm:2012gk}. In general, this $\Omega^{(2)}$-shift may transform the fluxes in curved indices in a non-trivial way, see e.g.~(3.10) of~\cite{Borsato:2021gma}.}. In addition, we may also transform the twist $U$ by an  automorphism $W$ of $\mathfrak{d}$ as $U'_I{}^M = W_I{}^J U_J{}^M$, without transforming $\dot{\bm{E}}_{\bm A}{}^{\bm I}$.  
To have a SGT, this automorphism must be outer --- since an inner automorphism would instead correspond to the action of a finite generalised diffeomorphism \footnote{For an inner automorphism $W = \text{Ad}_g$ with $g\in D$ one can show that $W(U_I{}^M) = W_I{}^J U_J{}^M = e^{{\cal L}_x} (U_I{}^M) $ where ${\cal L}_x$ is the generalised Lie derivative along $x = x^I U_I$, where $g = e^{x^I T_I}$.} --- and it must satisfy $W_I{}^J\mathcal F_J=\mathcal F_I$.

\vspace{12pt}

\paragraph*{Deformation theory.}
When focusing on the  $\beta$-shifts in~\eqref{eqn:mastereq}, our SGTs are classified by \emph{trivial} deformations (i.e.~isomorphisms) of the double Lie algebra $\mathfrak d$ that induce \emph{non-trivial} deformations of the subalgebra $\mathfrak h$. Naively, one would therefore expect that the classification of SGTs is given by the Lie algebra cohomologies $H^m(\mathfrak h,\mathfrak h)$, i.e.~the equivalence classes of $m$-alternating maps with coefficients in $\mathfrak h$ that are cocycles but not coboundaries~\footnote{In general, infinitesimal deformations of $\mathfrak h$ are classified by $H^2(\mathfrak h,\mathfrak h)$. Because we consider Lie algebras over the real numbers, if $H^2(\mathfrak h,\mathfrak h)=0$ not only the infinitesimal deformations are trivial, but even their finite version will be trivial, and the algebra is said to be rigid. Examples of rigid Lie algebras are the simple algebras. If $H^2(\mathfrak h,\mathfrak h)\neq 0$, one can try to promote non-trivial infinitesimal deformations to finite ones by constructing the higher-order terms. Obstructions to construct the higher orders are classified by $H^3(\mathfrak h,\mathfrak h)$. See e.g.~\cite{10.2307/24902154}.}. However, a closer look at~\eqref{eqn:mastereq} reveals a simpler answer. As we are about to prove,  the classification is in fact dictated by the Lie algebra cohomologies $H^m(\mathfrak h,\mathbb R)$ of alternating maps with coefficients \emph{in the real numbers}.

In order to show that, let us introduce an index-free notation. 
We use the basis vectors/forms $\T^i$ and $\theta_i$ with the grading
$
  \deg \T^i = \begin{pmatrix} 1 & 0 \end{pmatrix}
$ and $
  \deg \theta_i = \begin{pmatrix} 0 & 1 \end{pmatrix} $,
which is required to define the $\wedge$-product
$
  a \wedge b = (-1)^{\deg a \cdot \deg b} b \wedge a $,
where $\deg a \cdot \deg b$ denotes the scalar product.
We will also use the exterior derivative
\begin{equation}
  \dd \theta_i = -\frac12 Q_i{}^{jk} \theta_j \wedge \theta_k
,\quad
\text{and}
\quad
  \dd \T^i = Q_j{}^{ki} \T^j \wedge\theta_k \,,
\end{equation}
that satisfies $\dd^2=0$ because the Jacobi identity for $Q_i{}^{jk}$ holds.
We also have the Leibniz rule
$  \dd ( a \wedge b ) = \dd a \wedge b + (-1)^{(\deg a)_2} a \wedge \dd b$
in combination with the $\wedge$-product. Note that $(\deg a)_2$ denotes the second component of the degree vector. Finally, we need the \emph{non-associative} hook product~\footnote{ If brackets are not given explicitly, we assume
$
  a \h b \h c \h d = ( ( a \h b ) \h c ) \h d 
$
and so on.}, which is defined by
\begin{equation}
  \T^i \h \theta_j = \delta^i{}_j,
\end{equation}
and extends by
$  a \h ( b \wedge c ) =  ( a \h b ) \wedge c - (-1)^{\deg a \cdot \deg b} b \wedge ( a \h c )$
to quantities of arbitrary degree. Moreover, note that the hook satisfies
$
  \dd ( a \h b ) = \dd a \h b - (-1)^{(\deg a)_2} \wedge \dd b
$.

After defining the quantities
\begin{equation}
\begin{aligned}
&  H = \tfrac16 H_{ijk} \T^i \wedge \T^j \wedge \T^k, \quad
 &   F = \tfrac12 F_{ij}{}^k \T^i \wedge \T^j \wedge \theta_k, \\
 & Q = \tfrac12 Q_i{}^{jk} \T^i \wedge \theta_i \wedge \theta_k, \quad
 & R = \tfrac16 R^{ijk} \theta_i \wedge \theta_j \wedge \theta_k ,
  \end{aligned}
\end{equation}
one checks that the equations
\begin{equation}\label{eq:Jacobi-2}
  \dd H = F \h F,\quad
  \dd F = 0 ,\quad
    \dd Q = 0 ,\quad
  H \h F = 0\,,
\end{equation}
are equivalent to the Jacobi identity of $\mathfrak d$ when assuming $R^{ijk} = 0$. 

Our condition~\eqref{eqn:mastereq} --- i.e. that the flux $R'$ must vanish after applying the SGT --- can be compactly rewritten in this language as
\begin{equation}\label{mastereq2}
  \boxed{  R' = \dd \beta - \tfrac12 F \h \beta \h \beta + \tfrac16 H \h \beta \h \beta \h \beta  = 0}\,,
\end{equation}
where $\beta=\tfrac12 \beta^{ij}\theta_i\wedge\theta_j$.
To analyse the above equation, let us  rewrite $\beta$ as an expansion  in a  deformation parameter $\zeta$, namely
\begin{equation}
  \beta = \sum_{n=1}^{\infty} \zeta^n \beta_n \,.
\end{equation}
From~\eqref{mastereq2}, it follows that $\beta_1$ is closed, $\dd \beta_1=0$. For $\beta_1$ exact, one can check that the corresponding infinitesimal transformation is trivial --- it is a generalised diffeomorphism up to  compensating $\rho$- and $b$-twists on $\bm{ \dot{E}_A{}^I}$ and $\bm{ U_I{}^M}$ that are inconsequential for $\bm{ {E}_A{}^I}$.
Therefore, the non-trivial solutions for $\beta_1$ are classified by the second Lie algebra cohomology $H^2(\mathfrak h,\mathbb R)$.
The higher-order contributions to the expansion of~\eqref{mastereq2} are all of the form $\dd\beta_n=z_n$ where $z_n$ are 3-forms that can be computed explicitly from~\eqref{mastereq2}. As a consequence of $\dd^2=0$, $z_n$ must be closed, a fact that in turn is ensured by the identities~\eqref{eq:Jacobi-2}. On the other hand, a solution for $\beta_n$ can be found only if $z_n$ is exact. Therefore, the obstructions to obtaining the higher-order expansions $\beta_n$ with $ n>1$ are given by  $H^3(\mathfrak h,\mathbb R)$. This concludes our discussion on the analysis of~\eqref{eqn:mastereq} and the related deformation theory.

\vspace{12pt}

\paragraph*{An example: the bi-YBWZ model and its deformations.} Let us apply our methods to a non-trivial example with non-vanishing $(F,Q,H)$-fluxes. We choose to work with the bi-Yang-Baxter-Wess-Zumino (bi-YBWZ) model, which is known to be integrable on the worldsheet~\footnote{Here we want to consider SGTs of integrable $\sigma$-models, and as such we do not need to consider ${\cal F}_I$.}. This integrable deformation was constructed in \cite{Delduc:2017fib} and formulated as an ${\cal E}$-model (a first order doubled worldsheet model) in \cite{Klimcik:2020fhs}. From \cite{Klimcik:2020fhs} the doubled Lie group is known to be $D = G^{\mathbb{C}}$ for $G$ a real, semi-simple Lie group. The maximally isotropic subalgebra $\mathfrak{h}$ generated by $\tilde{T}^i$ is 
\begin{equation}
\mathfrak{h} = \left\{ \frac{e^{-i \rho_l} - e^{-\rho_l R}}{\sin\rho_l} y \ | \ y\in \text{Lie}(G) \right\} ,
\end{equation}
where $\rho_l \in ] -\pi , \pi [ $ is the  parameter denoting the deformation of the left symmetry group. The remaining generators of $\mathfrak{d}$ are those of $\text{Lie}(G)$ denoted by $\tilde{T}_i = t_i$. Here we are using tildes for $\tilde{T}_I = (\tilde{T}^i , \tilde{T}_i )$ because in this basis the metric $\tilde{\eta}_{IJ}$ is not canonical and has more generally the  form~\eqref{eq:noncanonicalmetric}. It is given by
\begin{equation}
  \tilde{\eta}_{IJ} = \frac{4\kappa}{\sin\rho_l} \mathrm{Im} \, \mathrm{tr} \left( e^{i\rho_l} \tilde{T}_I \tilde{T}_J \right)  ,
\end{equation}
with $\kappa$ the  WZ level. While the full structure constants ${\tilde{\cal F}}_{IJ}{}^K$  have both $\tilde{\cal F}^{ijk} = 0$ and $\tilde{\cal F}_{ijk} = 0$, due to the non-canonical basis the generalised fluxes $\tilde{\cal F}_{IJK} = \tilde{\cal F}_{IJ}{}^L \tilde{\eta}_{LK}$ have in fact only $\tilde{R}^{ijk} = 0$. The remaining fluxes are
$\tilde{H}_{ijk} = \tilde{F}_{[ij}{}^l \tilde{\eta}_{k]l}$, $\tilde{F}_{ij}{}^k = \tilde{\cal F}_{ij}{}^k$ and $\tilde{Q}^{ij}{}_k = \tilde{\cal F}^{ij}{}_k$.
For brevity, we omit the expressions for the fluxes in the canonical basis obtained by applying \eqref{eq:transformationcanonical} (notice that $R^{ijk}$ remains vanishing). 

We can use our previous discussion on the deformation theory to identify possible deformations of this model. Choosing $G=SU(2)$, for example, one finds that the corresponding subalgebra $\mathfrak h$ of $\mathfrak{d}$ has trivial second cohomology, so that there exists no non-trivial deformation. On the other hand, taking for example $G=SU(2)_L\times SU(2)_R$, one finds that $H^2(\mathfrak h,\mathbb R)$ is given by $\bar\theta_1\wedge \bar\theta_2$ where $\bar\theta_{1,2}$ are the two Cartan generators of $\mathfrak h$. Moreover, because in this case $H^3(\mathfrak h,\mathbb R)=0$, there is no obstruction to complete this to a finite deformation,  and in fact it turns out that the infinitesimal one is also the finite solution of \eqref{eqn:mastereq} (i.e.~the expansion truncates at first order).

For completeness, let us mention that the generalised fluxes in the canonical basis provide all the information needed  to construct the twist $U_I{}^M$ from \eqref{eq:canonicaltwist}. However, to construct the full  model, we also need an expression for $\dot{{E}}_{A}{}^{I}$ which --- because of the absence of spectators $(D = d)$ --- is a constant matrix. From the relation between ${\cal E}$-models and DFT laid out in \cite{Demulder:2018lmj} it is known that $\dot{{E}}_{A}{}^{I}$ can be obtained from the ${\cal E}$-operator as
$\dot{\mathcal H}^{IJ} = \dot{{E}}_{A}{}^{I} \eta^{AB} \dot{{E}}_{B}{}^{J} = \eta^{IL} {\cal E}_L{}^J $. It can be extracted from its $\pm 1$-eigenspaces ${\cal E}_\pm$ given by \cite{Klimcik:2020fhs} 
\begin{equation}
{\cal E}_\pm = \left\{ \left( \alpha^{\pm 1} - e^{-i \rho_l} e^{-\rho_r R} \right)x \ | \ x \in \text{Lie}(G) \right\} ,
\end{equation}
where $\rho_r \in ]-\pi , \pi [$ is the parameter transforming the right symmetry group and $\alpha \in ]-1,1[$ is an additional parameter. They both enter the model when expressing  the matrix ${\cal E}_I{}^J$ in the canonical basis.

\vspace{12pt}

\paragraph*{Conclusions.}
In this Letter we have given an algebraic classification of SGTs in (modified) SUGRA, as well as of canonical transformations of 2-dimensional $\sigma$-models. Our classification is complete for the families of SGTs that are compatible with a gSS ansatz in target space. 
Our results cover the homogeneous Yang-Baxter deformations  (when only $F_{ij}{}^k\neq 0$, and $\beta\neq 0, n=0$), non-abelian and Poisson-Lie T-duality (respectively when only $F_{ij}{}^k\neq 0$ or both $F_{ij}{}^k\neq 0, Q_i{}^{jk}\neq 0$, and $\beta=0,n=1$), the deformed T-dual models of~\cite{Borsato:2016pas,Borsato:2018idb} (when only $Q_{i}{}^{jk}\neq 0$ and $\beta\neq 0$), as well as deformations of Poisson-Lie symmetric backgrounds \cite{Borsato:2021gma} (when $F_{ij}{}^k\neq 0,Q_{i}{}^{jk}\neq 0$ and $\beta\neq 0$). The present results complete the classification of SGTs initiated in~\cite{Borsato:2021gma}, in particular clarifying and generalising the SGTs involving $H_{ijk}\neq 0$.

Starting from the master equation~\eqref{eqn:mastereq}, we have shown that the deformations are classified by the Lie algebra cohomologies $H^2(\mathfrak h,\mathbb R)$ and $H^3(\mathfrak h,\mathbb R)$ of the maximally isotropic subalgebra $\mathfrak h$ of the double Lie algebra $\mathfrak d$.

Let us emphasize that, when having in mind integrable $\sigma$-models, our results are valid also when considering the more recent formulations in terms of ${\cal E}$-models~\cite{Klimcik:2015gba,Klimcik:2016rov} or 4-dimensional Chern-Simons theory~\cite{Costello:2019tri,Vicedo:2019dej}, where the construction requires the choice of a maximally isotropic subgroup of a Drinfel'd double.

We have also discussed in detail the relation to modified SUGRA, and the conditions under which the SGTs remain in the realm of standard supergravity.
While in this Letter we have focused on the NSNS sector of supergravity, it is possible to include also the RR fields of type II. To do that, it is enough to assume that, together with the NSNS fields, also the RR ones are compatible with a gSS ansatz. This requirement then allows one to construct a spinor $\ket{\mathfrak F}$ that encodes the information on RR fields and that remains invariant under the SGTs. This invariance can then be used to derive the corresponding transformation rules of RR fields. For more details we refer to section 4 of~\cite{Borsato:2021gma}, where the same conventions of the present paper are used. 

We believe that our results will be useful to construct new backgrounds with motivations in the AdS/CFT correspondence, as well as new integrable models. It is intriguing to notice that the two moduli spaces --- of supergravity solutions on one side and of integrable models on the other --- may be (at least partially) organised in terms of the same algebraic structure. It would be interesting to understand whether there is a deeper relation than this. 

It would be also exciting to extend our results to the exceptional case, as it would lead to a classification of SGTs in string theory that involve U-duality transformations, rather than only T-duality. This study would go beyond the recent proposals of  U-duality generalisations of Poisson-Lie duality and deformations of e.g.~\cite{Sakatani:2019zrs,Malek:2019xrf}.

\vspace{12pt}

\begin{acknowledgments}
\paragraph*{Acknowledgements:}
We thank  Giacomo Piccinini for useful discussions. RB and SD are supported by the fellowship of ``la Caixa Foundation'' (ID 100010434) with code LCF/BQ/PI19/11690019,
by AEI-Spain (FPA2017-84436-P and Unidad de Excelencia Mar\'\i a de Maetzu MDM-2016-0692), by Xunta de Galicia-Conseller\'\i a de Educaci\'on (Centro singular de investigaci\'on  de  Galicia  accreditation  2019-2022), and  by the European Union FEDER. 
\end{acknowledgments}

\bibliographystyle{apsrev4-1}
\bibliography{literature}

\end{document}